\documentclass[12pt,preprint]{aastex}
\usepackage{natbib}

\shorttitle{Mass of the Black Hole in NGC 4151}
\shortauthors{Metzroth, Onken, \& Peterson}

\begin{document}

\title{The Mass of the Central Black Hole in the Seyfert Galaxy 
NGC~4151}

\author{Kyle G.~Metzroth\altaffilmark{1,2},
Christopher A.~Onken\altaffilmark{1,3}, 
and 
Bradley M.~Peterson\altaffilmark{1}}
\altaffiltext{1}{Department of Astronomy, The Ohio State University, 
Columbus, OH 43210; peterson@astronomy.ohio-state.edu}
\altaffiltext{2}{Current address: Department of Mechanical Engineering,
The Ohio State University, 650 Ackerman Road, Columbus, OH 43202;
metzroth.1@osu.edu}
\altaffiltext{3}{Current address: NRC Herzberg Institute of 
Astrophysics, 5071 West Saanich Road, Victoria, BC V9E 2E7, CANADA;
Christopher.Onken@nrc-cnrc.gc.ca}

\begin{abstract}
In order to improve the reverberation-mapping based
estimate of the mass of the central supermassive black hole
in the Seyfert 1 galaxy NGC~4151, we have 
reanalyzed archival ultraviolet monitoring spectra from
two campaigns undertaken with the {\em International
Ultraviolet Explorer}. We measure emission-line time delays
for four lines, \ion{C}{4}\,$\lambda1549$,
\ion{He}{2}\,$\lambda1640$,
\ion{C}{3}]\,$\lambda1909$, and
\ion{Mg}{2}\,$\lambda2798$, from both campaigns. We combine
these measurements with the dispersion of the variable part
of each respective emission line to obtain the mass of the central
object. Despite the problematic nature of some of the
data, we are able to measure a mass of 
$(4.14 \pm 0.73) \times 10^7\,M_\odot$, although this,
like all reverberation-based masses, is probably systematically
uncertain by a factor of 3--4.
\end{abstract}

\section{INTRODUCTION}

Reverberation mapping (Blandford \& McKee 1982; Peterson 1993) of
active galactic nuclei (AGNs) is
used to characterize the size of the broad-line region (BLR) in
these objects by measuring the time delay between continuum changes
and the response of the emission lines.
By combining the reverberation time delay, or ``lag,'' with the
width of the variable part of the emission line, it is possible
to estimate the mass of the central object, presumably a 
supermassive black hole, under the assumption that the dynamics
of the BLR gas are dominated by gravity. In this case, the mass
is given by
\begin{equation}
M_{\rm BH} = \frac{f c \tau \Delta V^2}{G},
\end{equation}
where the size of the BLR is given by the light-travel time
$c\tau$ and $\tau$ is the emission-line time delay,
$\Delta V$ is the width of the emission line, $G$ is
the gravitational constant, and $f$ is a factor of
order unity that depends on the geometry, kinematics,
and inclination of the BLR. 
Two lines of evidence argue that the reverberation-based
mass estimates have some veracity:
\begin{enumerate}
\item Different emission lines have different response times,
and these are inversely correlated with line width in a manner
consistent with a virialized BLR, i.e., $\tau \propto 
\Delta V^{-2}$ (Peterson \& Wandel 1999, 2000; Onken \& Peterson 2002;
Kollatschny 2003). Moreover, at least in the 
particularly well-studied case of 
the H$\beta$ line in NGC~5548, the lag and line width change
over time in response to luminosity changes, and the virial
relationship seems to be preserved (Peterson et al.\ 2004;
Cackett \& Horne 2006). 
\item There is a relationship between the reverberation-based
black hole mass $M_{\rm BH}$ and host-galaxy bulge velocity dispersion
$\sigma_*$ that is consistent with the same correlation,
the $M_{\rm BH}$--$\sigma_*$ relationship, that is observed
in quiescent galaxies (Ferrarese \& Merritt 2000;
Gebhardt et al.\ 2000a,b; Ferrarese et al.\ 2001;
Tremaine et al.\ 2002; Onken et al.\ 2003, 2004; Nelson et al.\ 2004).
\end{enumerate}
With respect to the second point, the consistency of the
$M_{\rm BH}$--$\sigma_*$ relationship between AGNs and
quiescent galaxies allows us to calibrate the reverberation-based
mass scale to that of quiescent galaxies by determining
a statistical mean value for the scaling constant $f$ in
eq.\ (1), as was done by Onken et al.\ (2004).

Reverberation results also show 
that there is a simple relationship between
the size of the BLR $R = c\tau$ and luminosity $L$ of the form
$R \propto L^{\alpha}$, where $\alpha \approx 0.5$, but depends
somewhat on the luminosity measure and also, presumably, 
the particular emission line for which $R$ is measured
(Kaspi et al.\ 2000, 2005; Bentz et al.\ 2006a). This is an
especially important result, since the mass of the black hole
in any AGN can then be estimated through a single measurement
of luminosity and line width, thus enabling mass estimates
for large populations of AGNs (Wandel, Peterson, \& Malkan 1999; 
Vestergaard 2002, 2004; McLure \&
Jarvis 2002; Kollmeier et al.\ 2006; Vestergaard \& Peterson 2006).

Reverberation mapping is currently the only 
broadly applicable method by
which we can directly measure AGN black hole masses and it holds
future promise because it is the only direct method of black hole mass
measurement that does not depend on angular resolution. Moreover,
reverberation-based mass measurements anchor the 
calibration for masses
based on radius--luminosity scaling relationships. 
Thus, given the importance
of the reverberation results, we have undertaken a variety
of programs designed to improve the reverberation measurements from
existing data, in addition to carrying out new reverberation-mapping
experiments. These efforts have included compilation 
and consistent reanalysis of most existing reverberation data
(Peterson et al.\ 2004). In the particular case of NGC~3783, we
completely remeasured and reanalyzed the data obtained with the
{\em International Ultraviolet Explorer (IUE)} using improved
spectral extractions (Onken \& Peterson 2002), which resulted
in a remarkable improvement in the precision of the central black
hole mass. In this contribution, we undertake a similar reanalysis
on the {\em IUE} spectra of NGC~4151, motivated at least in
part by the fact that NGC~4151 is one of the few AGNs in which
measurement of the black hole mass by other means is plausible,
which would thus enable a direct comparison between masses measured
by reverberation and those measured by other direct methods.

In this contribution, we re-examine spectra from two 
ultraviolet monitoring campaigns undertaken
with {\em IUE} in 1988 (Clavel et al.\ 1990) 
and in 1991 (Ulrich \& Horne 1996), and compare the results with
those from two ground-based optical monitoring programs that
were reanalyzed by Peterson et al.\ (2004). A third ultraviolet
monitoring program on NGC~4151 in 1993 (Crenshaw et al.\ 1996)
is not revisited here because the program was too short (9.3 days) 
to yield meaningful results on the emission-line responses.
In section 2, we describe the processing and measurement of the
ultraviolet spectra. Our reverberation analysis is described in
section 3, and in section 4 we discuss our measurement of the
black hole mass. Our conclusions appear in section 5.

\section{THE ULTRAVIOLET SPECTRA}
The databases for both {\em IUE} monitoring programs consist
of multiple observations made with the Short Wavelength
Prime (SWP) and Long Wavelength Prime (LWP) 
cameras in the low-dispersion mode with a 
large ($10''\times20''$) oval  aperture.
The SWP spectra cover the wavelength range 1150\,\AA\ to 2000\,\AA, 
while the LWP spectra cover the range 1800\,\AA\ to 3300\,\AA,
although the LWP spectra are nearly worthless shortward of 
about 2200\,\AA.
The first data set that we examine was obtained during the
period 1988 November 29 to 1989 January 30 and is comprised of
33 SWP and 22 LWP spectra (Clavel et al.\ 1990).
The second set was obtained between
1991 November 9 and December 15, and is comprised of 44 SWP and 
37 LWP spectra (Ulrich \& Horne 1996).
The original spectral images were processed with
NEW Spectral Image Processing System
(NEWSIPS), replacing the older {\em IUE} Spectral Image
Processing System (IUESIPS) extractions that were used in the
original analysis. Compared to the IUESIPS spectra,
the NEWSIPS spectra have improved
photometric accuracy and higher signal-to-noise ratio ($S/N$);
AGN spectra processed with NEWSIPS show a 10\%--50\% increase
in $S/N$ (Onken \& Peterson 2002 and references therein).

Measurements were made of each of the SWP and LWP
spectra. The continuum flux was measured in the SWP spectra
over a 30\,\AA\ bandpass centered on 1355\,\AA\ in the
observed frame. Emission-line fluxes were measured by 
defining nominally line-free regions bracketing the lines,
fitting a linear continuum between these regions and measuring
the flux above the interpolated continuum. The wavelengths of
the continuum fitting regions and the limits of the line integration
are given for four emission lines, 
\ion{C}{4}\,$\lambda1549$,
\ion{He}{2}\,$\lambda1640$,
\ion{C}{3}]\,$\lambda1909$, and
\ion{Mg}{2}\,$\lambda2798$,
in Table 1. We did not include measurements of the
Ly$\alpha\,\lambda1215$+\ion{N}{5}\,$\lambda1240$ complex because
this spectral region is hopelessly contaminated by geocoronal
Ly$\alpha$ emission. We also attempted to measure the flux
in the \ion{Si}{4}$\,\lambda1400$ feature, but the results were
very poor because the line is so weak. These measurements were
discarded. 
There were a few spectra in which the measured fluxes deviated strongly
from more plausible values obtained from redundant 
spectra obtained at the same epoch (i.e., during the same 8-hour 
observing
shift).  We are not always able to identify
specific causes of these deviant points; some effects,
such as grazing-incidence cosmic rays, are
notoriously difficult to remove from {\em IUE} spectra through the 
standard 
pipeline processing methods such as NEWSIPS.  We elected to simply 
remove 
the strongly deviant values from the light curves before processing.  
The measurements of the continuum and emission-line
fluxes used in this analysis are given in Tables 2--5.

Multiple measurements that were obtained within a single 8-hour
{\em IUE} observing shift were compared to determine uncertainties
in the fluxes on the assumption that no real detectable variability
occurred on such short time scales. Following this, data points 
obtained
in a single shift were replaced by a weighted average to form
the final light curves shown in the left-hand columns of 
Figs.\ 1--2 and which were used as the
basis for the time-series analysis described in \S{3.1}.
The statistical properties of these light curves are summarized in
Table 6. Column (1) identifies the spectral feature and column (2)
gives the total number of measurements in the time series. The 
average and median intervals between individual data points are
given in columns (3) and (4), respectively. The mean flux and
its standard deviation appear in column (5). The mean fractional
error, based on comparison of closely spaced observations, is
given in column (6). Column (7) gives the ``excess variance'' for
each light curve, computed as
\begin{equation}
\label{eq:fvar}
F_{\rm var}=\frac{\sqrt{\sigma^{2} - \delta^{2}}}{\langle f \rangle},
\end{equation}
where $\sigma^{2} $ is the variance of all of the fluxes, $\delta^{2} $ 
is
the mean square uncertainty of the fluxes and $\langle f \rangle$ is 
the mean
flux for all observations. Also listed in column (8) is $R_{\rm max}$, 
the ratio of the maximum and minimum fluxes for each time series.

\section{DATA ANALYSIS}
\subsection{Time Series Analysis}

To find the time delay between the continuum and emission-line
variations, we cross-correlate each of the emission-line light curves 
with that of the 1355\,\AA\ continuum. The methodology we employ is the 
interpolation correlation function method as described by White \& 
Peterson (1994). The cross-correlation functions (CCFs) for the 
emission lines are shown in the right-hand columns of 
Figs.\ 1 and 2. In order to assess uncertainties in the time-delay 
measurements, we employ the model-independent Monte Carlo FR/RSS method 
described by Peterson et al.\ (1998), with some modifications 
introduced by Peterson et al.\ (2004), which works as follows. For a 
single realization, a light curve of $N$ data points is sampled $N$ 
times without regard to whether or not any given point has been 
previously selected; this is called
``random subset sampling,'' or RSS. Any 
data point that is selected $M$ times has its uncertainty 
in flux reduced by a 
factor $M^{1/2}$. The fluxes in each of the selected $N$ points are 
altered by random Gaussian deviates based on their adopted error bars; 
we refer to this as ``flux randomization,'' or FR. The subset of these 
points, sampled and altered by the FR/RSS algorithm, are then 
cross-correlated as though they were real data. This yields a CCF like 
those seen in Figs.\ 1 and 2. We locate the peak value $r_{\rm max}$ of  
the CCF, which occurs at a time lag $\tau_{\rm peak}$. We 
also compute the centroid of the CCF, $\tau_{\rm cent}$, based on those 
points near the peak with values $r \ga 0.8r_{\rm max}$. A large 
number of such Monte Carlo realizations builds up a cross-correlation 
peak distribution (CCPD) for $\tau_{\rm peak}$ and a
cross-correlation centroid distribution (CCCD) for $\tau_{\rm cent}$. 
As 
argued elsewhere, $\tau_{\rm cent}$ is more repeatable and has a 
clearer physical interpretation, so we prefer it to $\tau_{\rm peak}$. 
We thus take the average values of the CCCD and CCPD to be  
$\tau_{\rm cent}$ and $\tau_{\rm peak}$, respectively.
The uncertainties $\Delta \tau_{\rm upper}$ and $\Delta \tau_{\rm 
lower}$ are computed such that 15.87\% of the CCCD realizations have
values $\tau < \tau_{\rm cent} - \Delta \tau_{\rm lower}$ and 
15.87\% of the CCCD realizations have
values $\tau > \tau_{\rm cent} + \Delta \tau_{\rm upper}$, with
the errors in $\tau_{\rm peak}$ defined similarly. These uncertainties
correspond to $\pm 1 \sigma$ for a Gaussian distribution.
The values so computed for these 
data sets are given in Table 7. Note that the errors are generally 
asymmetric, but usually not strongly so. It should be noted that 
formally these observed-frame measurements need to be corrected for
time dilation by division by $(1 + z)$, where the systemic redshift of 
NGC~4151 is $z = 0.00332$.

\subsection{Line Width Measurement}
To obtain the black hole mass, we also need to measure the width of 
each emission line. Indeed, we wish to measure the 
line-of-sight velocity distribution for the {\em variable} part of the 
emission-line, specifically avoiding contaminating non-variable (on 
reverberation timescales) components, such as a contribution from the 
much larger narrow-line region. We use all of the spectra obtained 
during the observing campaign to construct a mean spectrum and a 
root-mean-square (rms) spectrum; the rms spectrum isolates the variable 
part of the emission lines. In Figs.\ 3 and 4, we show the mean and rms 
spectra for the SWP and LWP images from 1988 and 1991, respectively.

To measure the emission-line widths, we first interpolate the rms 
continuum
under the emission lines by fitting a linear continuum
in the continuum windows given in Table 1. We then characterize the 
line width in two ways, by its full-width at half maximum (FWHM) and by 
the second moment of the line profile, i.e., the line dispersion 
$\sigma_{\rm line}$, as described by Peterson et al.\ (2004). To 
evaluate the uncertainties in these measurements, we follow the 
procedure described by Peterson et al.\ (2004), using a procedure 
similar in spirit to that used to evaluate uncertainties in the time 
delays. For a sample of $N$ spectra, we select $N$ spectra at random, 
in each case
without regard to whether or not a particular spectrum has been 
previously selected. 
These $N$ random spectra are used to construct mean and rms spectra, 
and both line width measurements are made for each emission line. This 
constitutes one Monte Carlo realization. A large number of similar 
realizations yields a mean and standard deviation for each 
line-width measure. 
Line width measurements and associated uncertainties
for each of the emission lines are given in 
Table 8. The emission-line widths here have been converted
to the rest-frame of NGC~4151 and have been
adjusted as described by Peterson et al.\ (2004) to account for
the resolution
of the {\em IUE} spectrograph. We note that we prefer $\sigma_{\rm 
line}$ over FWHM as a line width measure for a variety of reasons, 
including that it is generally more 
repeatable in noisy spectra (which rms spectra often are).

Two obvious difficulties are apparent upon inspection of Figs.\ 3 and 
4. The first of these is that the core of the \ion{C}{4} emission line 
is strongly self-absorbed. This is very apparent in the rms spectra, 
and is especially strong in the 1991 data. It is impossible to correct 
for this absorption since we do not know the intrinsic unabsorbed 
\ion{C}{4} profile. However, we can try to assess the severity of the 
systematic uncertainty by modeling the intrinsic profile in a variety 
of ways to see how much the line width measurement might change. We 
experimented with various means of accounting for the effects of the 
absorption, and the largest change in line width was obtained by 
fitting the unabsorbed wings of the emission line with a Gaussian. The 
Gaussian so constructed had a value of $\sigma_{\rm line}$ that was 
about 10\% smaller than that obtained in our direct measurement,
and the value of FWHM, which is more sensitive to the flux at line
center, decreased by more than 20\%. 
We can regard our measurement of the \ion{C}{4} line 
width as an upper limit that is probably not a terrible overestimate of 
the true unabsorbed value of $\sigma_{\rm line}$.
In any case, the effect on the black hole mass estimate is
quite insignificant given the level of
accuracy that we can currently achieve in black hole mass measurement.
We also note that the \ion{Mg}{2} line is self-absorbed, but the
absorption is much weaker and the emission line in the rms spectrum
is too noisy for the absorption feature to be apparent.

A second difficulty is that the \ion{C}{4} and \ion{He}{2} emission 
lines are strongly blended in their wings. Since these cannot be
uniquely deblended, we make the approximation that the both
lines are approximately symmetric about line center and use the 
unblended
half of each line (the short wavelength side for \ion{C}{4} and
the long wavelength side for \ion{He}{2}) to determine the width.
These are the values of the line widths that appear in Table 8.

\section{THE BLACK HOLE MASS}

As noted earlier and as observed in other sources, we expect
that if the dynamics of the BLR are dominated by the gravitational
force of the central black hole, we should see a virial relationship
between emission-line widths and time lags of the form
$\Delta V \propto \tau^{-1/2}$. To test this for NGC~4151, 
we plot the emission-line line widths from the rms spectra (Table 8)
versus the measured time delays (Table 7) in Fig.\ 5.
We supplement the UV data with measurements of the
H$\alpha$ and H$\beta$ line widths and lags from
Peterson et al.\ (2004), based on data originally published
by Maoz et al.\ (1991) and Kaspi et al.\ (1996). 
There is considerable scatter in Fig.\ 5, but with the
exception of the H$\alpha$ and H$\beta$ measurements
based on the Kaspi et al.\ (1996) monitoring program,
the scatter is similar to that seen in other sources
(cf.\ Peterson et al. 2004). The range of lags in
this diagram is less than a factor of 3, whereas in
the similar plot for NGC~5548 (cf.\ Fig.\ 3 of Peterson et al.\
2004), the range of lags is nearly a factor of 15.
Moreover, given the limited quality of
the monitoring data on NGC~4151, it is perhaps surprising that
the results are as good as they appear to be. Neither of the
two UV monitoring data sets on NGC~4151 are remarkably good: 
the 1988 {\em IUE} data
are slightly undersampled and the duration of the 1991 experiment
was somewhat short, especially in the case
of the \ion{Mg}{2} observations,
the effect of which is apparent in 
its flat-topped CCF (Fig.\ 2).
The only UV line not affected by self-absorption or blending
is \ion{C}{3}], for which the variations are comparatively weak.

The two existing optical data sets are even more problematic.
In the case of the 1988 data from Maoz et al.\ (1991),
the emission-line lags appear to be well determined, but
it is difficult to measure 
reliably the width of the broad-line component in these spectra.
In NGC 4151, the narrow-line components are much
stronger relative to the broad-line components
than they are in most type 1 AGNs. The spectra from this campaign
are rather low resolution, so the 
[\ion{O}{3}] lines are partially blended with one another.
Moreover, the line-spread
function appears to vary among the spectra, possibly as a result
of drift in the large aperture that was used to ensure an accurate
flux calibration. The combination of these factors makes it hard
to isolate the broad-line component and
determine their  widths with great confidence.
Nevertheless, the H$\alpha$ and H$\beta$ lags and line widths
as plotted in Fig.\ 5 are reasonably
consistent with the virial relationship derived from the UV lines
in this object.

On the other hand, the optical data described by Kaspi et al.\ (1996)
present some serious difficulties for a virial interpretation.
However, in this particular case, the nature
of the variations during this campaign were not favorable for
reverberation analysis --- both the continuum and emission lines
showed nearly monotonically increasing flux throughout the monitoring
period, and the amplitude of variation was 
relatively low. Without a strong change in the
sign of the first derivative of the light curves, as seen
in the light curves in Figs.\ 1 and 2, it is difficult to
obtain a highly reliable reverberation lag. In an attempt
to mitigate the unfavorable effects of a monotonic rise,
we experimented with removing the long-term trend
in these data prior to cross-correlating the time series
(i.e., ``detrending'' the data, cf.\ Welsh 1999), and
this had the effect of moving the already-small lag even closer to zero. 
Our suspicion is that this lag measurement is spurious.
It seems likely that at such a low level of variability,
there are correlated errors between the continuum and emission-line
measurements, which manifest themselves as a correlated
signal at zero lag. We have not been able to demonstrate
this conclusively and thus need to keep in mind the possibility
that these measurements represent an actual deviation from
the virial relationship. However, given the unfavorable nature of
the observed variations, we are more inclined at the
present time to simply disregard this particular data set.

The best-fit slope to all of the data points in Fig.\ 5 
is $-1.52 \pm 0.84$, which differs from the virial slope
of $-0.5$ by only $1.2\sigma$. Obviously additional, better
data will be required to determine whether or not the 
virial relationship between lag and line widths holds in
the case of NGC~4151.

Setting aside this difficulty, we nevertheless proceed with
an estimate of the mass of the central black hole by
using eq.\ (1)
with the scaling factor $f = 5.5$, as determined by Onken et al.\
(2004) by normalizing the AGN $M_{\rm BH}$--$\sigma_*$ to that for
quiescent galaxies. We use $\tau_{\rm cent}$ for the time delay and
$\sigma_{\rm line}$ to characterize the line width. If we use all of
the data in Fig.\ 5, we find a black hole mass
$M_{\rm BH} = (2.58 \pm 0.35) \times 10^{7}\,M_\odot$, based
on a weighted average of the individual mass estimates for each line.
If we restrict the mass estimate to the UV data points from the
present work, we obtain a mass estimate of 
$M_{\rm BH} = (8.55 \pm 1.26) \times 10^{7}\,M_\odot$.
Our preferred estimate is obtained by eliminating
the most problematic data, leaving only \ion{C}{3}],
\ion{He}{2}, and \ion{Mg}{2} from the two UV campaigns. 
The resulting mass estimate is
$M_{\rm BH} = (4.14 \pm 0.73) \times 10^{7}\,M_\odot$. We remind
the reader, however, that due to unquantified systematic
uncertainties (as embodied in the scale factor $f$ in eq.\ 1),
this is probably uncertain by a factor of 3--4
(Onken et al.\ 2004), as are all reverberation-based
mass estimates.

\section{CONCLUSIONS}

In this contribution, we have examined archival {\em IUE} spectra
that have been reprocessed with the NEWSIPS software. For two separate 
monitoring programs, we were able to obtain emission-line time delays
for \ion{C}{4}\,$\lambda1549$,
\ion{He}{2}\,$\lambda1640$,
\ion{C}{3}]\,$\lambda1909$, and
\ion{Mg}{2}\,$\lambda2798$. Unfortunately, the lags span a very
narrow range, even when optical Balmer line measurements from
ground-based campaigns are included, precluding a strong
test of the expected virial relationship, 
$\Delta V \propto \tau^{-1/2}$. Clearly additional data are
required to clarify the situation.

Ignoring this difficulty for the time being, we obtain an
estimate of the mass of the central black hole by
combining our time delay measurements with
line-width measurements. Based on the subset of lines that we regard as most 
reliable, we provide an estimate of the black hole mass of 
$M_{\rm BH} = (4.14 \pm 0.73) \times 10^{7}\,M_\odot$. This is a factor 
of $\sim3$ higher than the previous estimate of
$M_{\rm BH} = (1.33 \pm 0.46) \times 10^{7}\,M_\odot$ from
Peterson et al.\ (2004); the earlier estimate was based on the optical 
data from Kaspi et al.\ (1996), which are somewhat problematic.
Indeed, problems with the existing optical data have led us to 
carry out a new optical reverberation program, the results of
which will be reported elsewhere (Bentz et al.\ 2006b).

It is also worth noting that the higher mass estimate means
that the black hole radius of influence, $r = GM_{\rm BH}/\sigma_*^2 
\approx 19\,{\rm pc}$,
is larger than previously supposed, projecting to an angular
radius of $0.\!''28$. This makes NGC~4151 one of the best candidates
for black hole measurement by other direct means that depend on
high angular resolution.

\acknowledgments
We are grateful for support of this program by the
NSF through grant AST-0205964 
and by NASA through grant HST GO-09849 from the 
Space Telescope Science Institute.

\clearpage

\begin{figure}
\begin{center}
\epsscale{0.75}
\plotone{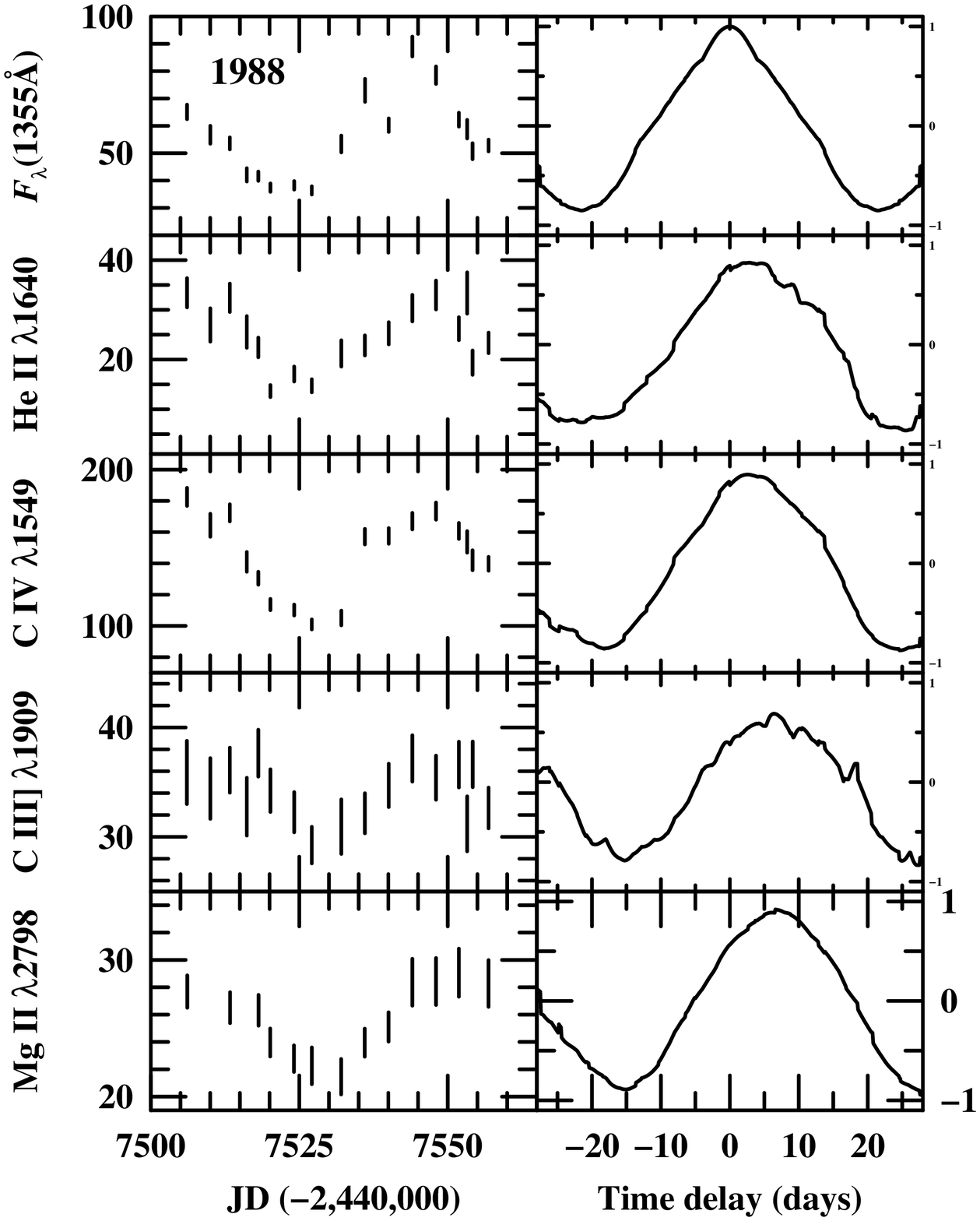}
\caption{Light curves and cross-correlation functions based on
the 1988 {\em IUE} observations of NGC~4151. The left column
shows flux as a function of Julian date, from top to bottom,
for the continuum flux at 1355\,\AA, and the emission-line
fluxes for \ion{He}{2}\,$\lambda1640$, \ion{C}{4}$\lambda1549$,
\ion{C}{3}]\,$\lambda1909$, and \ion{Mg}{2}\,$\lambda2798$,
as listed in Tables 2 and 3.
The continuum flux is plotted in units of
$10^{-15}$\,ergs\,s$^{-1}$\,cm$^{-2}$\,\AA$^{-1}$
and the emission-line fluxes are in units of 
$10^{-13}$\,ergs\,s$^{-1}$\,cm$^{-2}$.
The right column shows the result of cross-correlating
each of these light curves with the 1355\,\AA\ continuum
light curve; the top panel is thus the continuum autocorrelation
function.}
\end{center}
\end{figure}

\begin{figure}
\begin{center}
\epsscale{0.75}
\plotone{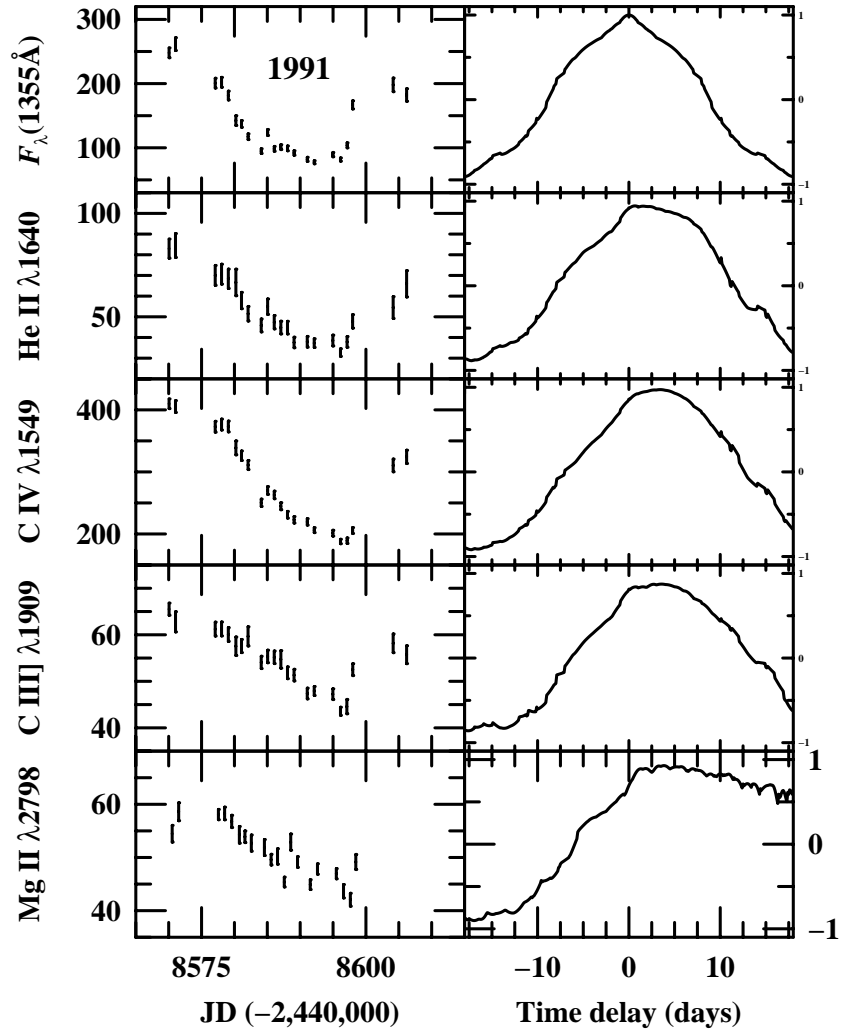}
\caption{Light curves and cross-correlation functions based on
the 1991 {\em IUE} observations of NGC~4151, as listed
in Tables 4 and 5, and plotted as in Fig.\ 1.}
\end{center}
\end{figure}

\begin{figure}
\begin{center}
\epsscale{1.0}
\plotone{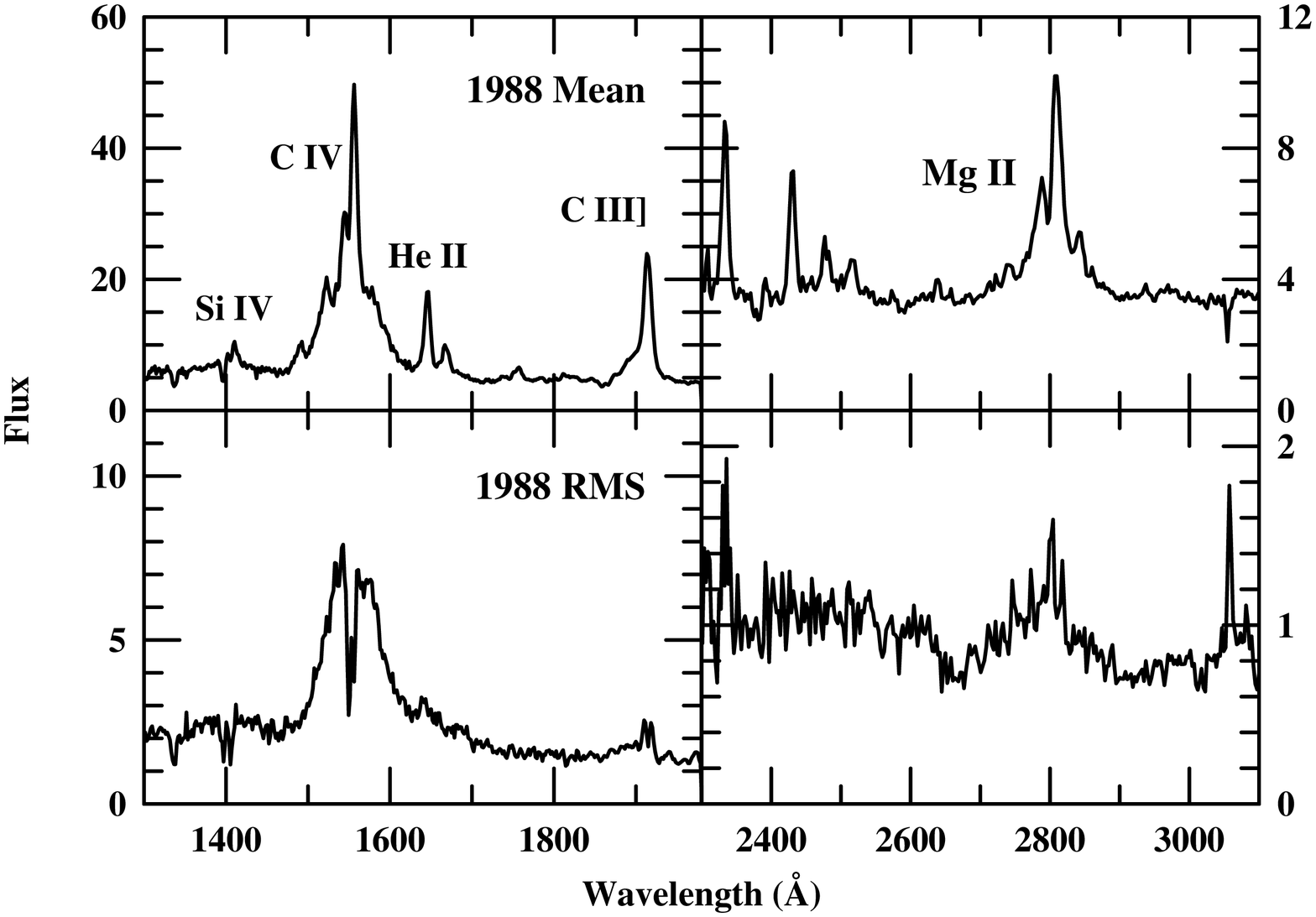}
\caption{Mean and rms spectra based on the 1988 {\em IUE}
observations of NGC~4151. The left column shows the
mean (upper panel) and rms (lower panel) spectra formed
from the SWP data.
The right column shows the mean (upper panel) and rms (lower panel) 
spectra formed from the LWP data. Fluxes are in units of
$10^{-15}$\,ergs\,s$^{-1}$\,cm$^{-2}$\,\AA$^{-1}$
and the spectra are plotted in the observed frame.}
\end{center}
\end{figure}

\begin{figure}
\begin{center}
\epsscale{1.0}
\plotone{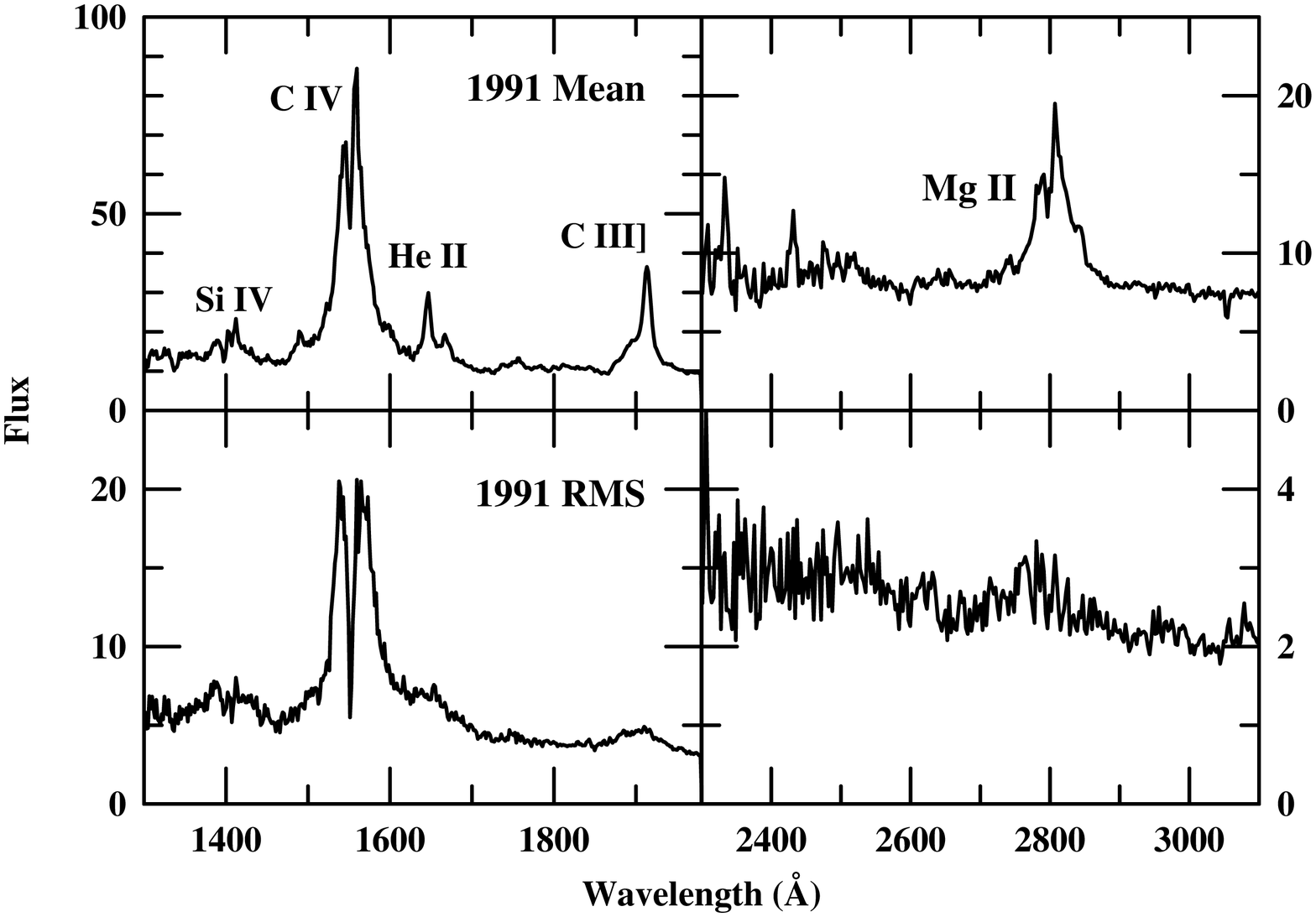}
\caption{Mean and rms spectra based on the 1991 {\em IUE}
observations of NGC~4151. The left column shows the
mean (upper panel) and rms (lower panel) spectra formed
from the SWP data.
The right column shows the mean (upper panel) and rms (lower panel) 
spectra formed from the LWP data. Fluxes are in units of
$10^{-15}$\,ergs\,s$^{-1}$\,cm$^{-2}$\,\AA$^{-1}$
and the spectra are plotted in the observed frame.}
\end{center}
\end{figure}

\begin{figure}
\begin{center}
\epsscale{1.0}
\plotone{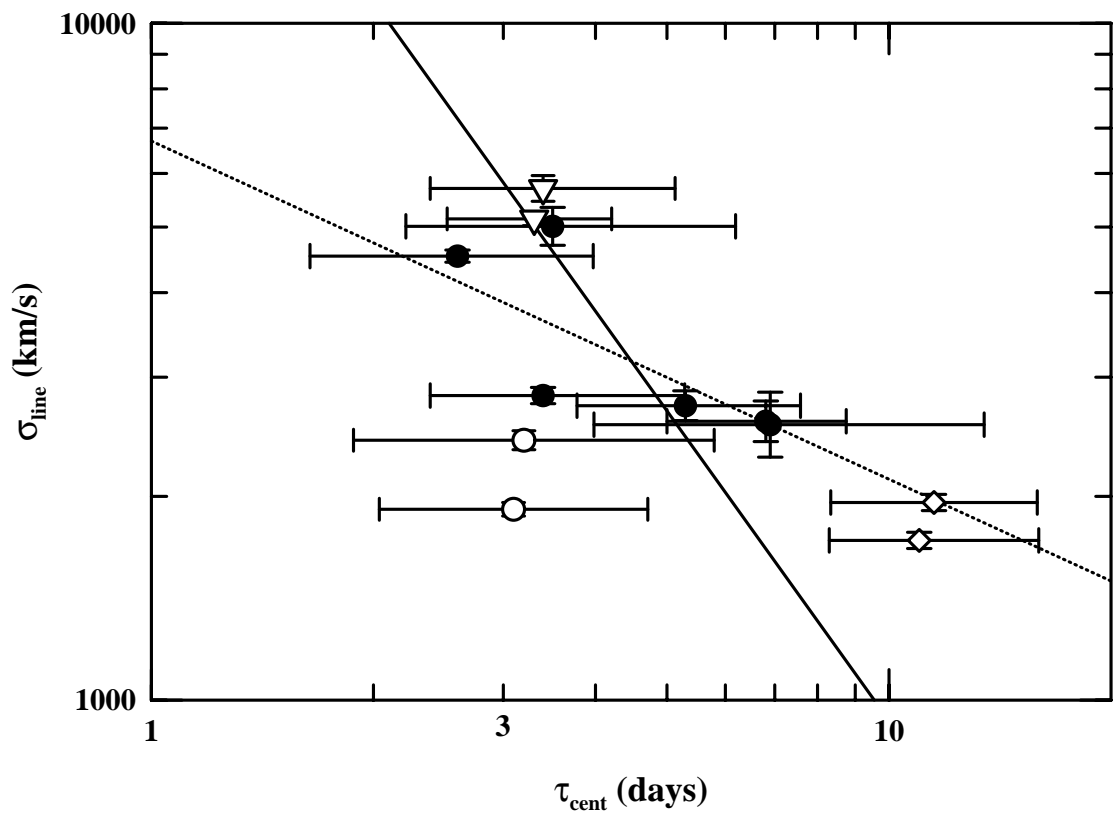}
\caption{Emission-line widths (as characterized by
line dispersion), versus time lags for lines in 
NGC\,4151. The closed circles are the UV emission
lines listed in Tables 7 and 8, except for the
\ion{C}{4} lines, which are shown as inverted
open triangles. The open circles and diamonds are
hydrogen Balmer-line measurements from 1987 and 1993, respectively,
as analyzed by Peterson et al.\ (2004). The solid line
has a slope of $-1.52 \pm 0.84$ and 
is the best fit to all the data, and the dotted line is
the best fit for a forced slope of $-1/2$, i.e., a
virial relationship. The filled symbols represent the
more trustworthy measurements.}
\end{center}
\end{figure}

\clearpage

%%% TABLE ! %%%
\begin{deluxetable}{lccc} 
\tablewidth{0pt} 
\tablecaption{Line Integration Limits} 
\tablehead{ 
\colhead{Emission} & 
\colhead{Line Integration} & 
\colhead{Continuum} & 
\colhead{Continuum}  \\ 
\colhead{Line } & 
\colhead{Limits (\AA)} & 
\colhead{Window (\AA)} & 
\colhead{Window (\AA)} 
} 
\startdata
1988 C\,{\sc iv}\,$\lambda1549$   & 1470--1620& 1440--1470& 1700--1730\\
1988 He\,{\sc ii}\,$\lambda1640$  & 1620--1700& 1465--1475& 1730--1740\\
1988 C\,{\sc iii}]\,$\lambda1909$ & 1863--1937& 1845--1855& 1965--1975 \\
1988 Mg\,{\sc ii}\,$\lambda2798$  & 2750--2853& 2600--2620& 2890--2900 \\
1991 C\,{\sc iv}\,$\lambda1549$   & 1475--1602& 1465--1475& 1700--1705 \\ 
1991 He\,{\sc ii}\,$\lambda1640$  & 1602--1700& 1465--1475& 1730--1740 \\ 
1991 C\,{\sc iii}]\,$\lambda1909$ & 1863--1937& 1845--1855& 1965--1975 \\ 
1991 Mg\,{\sc ii}\,$\lambda2798$  & 2750--2853& 2600--2620& 2890--2900  
\enddata
\end{deluxetable} 

%%% TABLE 2 %%%
\begin{deluxetable}{lccccc}
\tablewidth{0pt}
\tablecaption{1988 SWP Flux Measurements}
\tablehead{
\colhead{Image} &
\colhead{Julian Date} &
\colhead{ } &
\colhead{ } &
\colhead{ } &
\colhead{ } \\
\colhead{Name} &
\colhead{$-2,440,000$} &
\colhead{$F_{\lambda}$(1355\,\AA)\tablenotemark{1} }&
\colhead{C\,{\sc iv}\,$\lambda 1549$\tablenotemark{2}} &
\colhead{He\,{\sc ii}\,$\lambda 1640$\tablenotemark{2}} &
\colhead{C\,{\sc iii}]\,$\lambda 1909$\tablenotemark{2}} 
}
\startdata
SWP34845&7497.992 &\nodata             & 227.48 $\pm $ 10.20 & 53.17 $\pm $ 6.62& 34.56 $\pm $ 2.78 \\
SWP34868&7499.549 & 93.96 $\pm $ 5.42  & 229.65 $\pm $ 10.29 & 52.57 $\pm $ 6.54& 35.83 $\pm $ 2.89 \\
SWP34869&7499.630 & 98.90 $\pm $ 5.71  & \nodata             & 52.22 $\pm $ 6.50& \nodata            \\
SWP34998&7506.075 & 66.55 $\pm $ 3.84  & 184.00 $\pm $ 8.25 & 33.00 $\pm $ 4.11 & 35.89 $\pm $ 2.89 \\
SWP34999&7506.132 & 63.77 $\pm $ 3.68  & 181.19 $\pm $ 8.12 & 33.89 $\pm $ 4.22 & \nodata           \\
SWP35028&7510.017 & 56.74 $\pm $ 3.27  & 164.41 $\pm $ 7.37 & 26.95 $\pm $ 3.35 & 34.42 $\pm $ 2.77 \\
SWP35058&7513.273 & 55.10 $\pm $ 3.18  & 177.79 $\pm $ 7.97 & 31.98 $\pm $ 3.98 & 37.16 $\pm $ 2.99 \\
SWP35059&7513.335 & 52.24 $\pm $ 3.01  & 167.54 $\pm $ 7.51 & 32.88 $\pm $ 4.09 & 35.14 $\pm $ 2.83 \\
SWP35090&7516.165 & 42.08 $\pm $ 2.43  & 140.97 $\pm $ 6.32 & 25.55 $\pm $ 3.18 & 32.76 $\pm $ 2.64 \\
SWP35098&7518.066 & 39.73 $\pm $ 2.92  & 131.98 $\pm $ 5.92 & 24.39 $\pm $ 3.04 & 37.02 $\pm $ 2.98 \\
SWP35099&7518.135 & 43.86 $\pm $ 2.53  & 129.10 $\pm $ 5.79 & 20.99 $\pm $ 2.61 & 38.33 $\pm $ 3.09 \\
SWP35123&7520.111 & 37.91 $\pm $ 2.19  & 107.59 $\pm $ 4.82 & 12.30 $\pm $ 1.53 & 33.28 $\pm $ 2.68 \\
SWP35124&7520.166 & 36.93 $\pm $ 2.13  & 121.40 $\pm $ 5.44 & 15.95 $\pm $ 1.99 & 35.30 $\pm $ 2.84 \\
SWP35171&7524.109 & 38.22 $\pm $ 2.21  & 114.64 $\pm $ 5.14 & 17.60 $\pm $ 2.19 & 32.87 $\pm $ 2.65 \\
SWP35172&7524.171 & 38.29 $\pm $ 2.21  & 106.29 $\pm $ 4.76 & 16.64 $\pm $ 2.07 & 31.69 $\pm $ 2.55 \\
SWP35210&7527.075 & 33.30 $\pm $ 1.92  & 100.50 $\pm $ 4.51 & 13.06 $\pm $ 1.63 & 27.38 $\pm $ 2.21 \\
SWP35211&7527.143 & 41.09 $\pm $ 2.37  & 101.39 $\pm $ 4.55 & 17.84 $\pm $ 2.22 & 31.76 $\pm $ 2.56 \\
SWP35264&7532.083 & 53.37 $\pm $ 3.08  & 105.06 $\pm $ 4.71 & 21.22 $\pm $ 2.64 & 30.93 $\pm $ 2.49 \\
SWP35297&7536.019 & \nodata            & 153.93 $\pm $ 6.90 & 21.15 $\pm $ 2.63 & 31.92 $\pm $ 2.57 \\
SWP35298&7536.075 & 72.98 $\pm $ 4.21  & 160.55 $\pm $ 7.20 & 25.25 $\pm $ 3.14 & 32.40 $\pm $ 2.61 \\
SWP35330&7540.032 & 58.89 $\pm $ 3.40  & 160.04 $\pm $ 7.17 & 25.96 $\pm $ 3.23 & 36.94 $\pm $ 2.98 \\
SWP35331&7540.091 & 61.76 $\pm $ 3.56  & 155.24 $\pm $ 6.56 & 24.67 $\pm $ 3.07 & 32.93 $\pm $ 2.65 \\
SWP35374&7544.004 & 86.76 $\pm $ 5.00  & 163.48 $\pm $ 7.33 & 28.77 $\pm $ 3.58 & 39.64 $\pm $ 3.19 \\
SWP35375&7544.070 & 91.49 $\pm $ 5.28  & 170.93 $\pm $ 7.66 & 32.32 $\pm $ 4.02 & 35.22 $\pm $ 2.84 \\
SWP35388&7547.998 & 79.78 $\pm $ 4.60  & 175.48 $\pm $ 7.87 & 33.62 $\pm $ 4.18 & 34.70 $\pm $ 2.80 \\
SWP35389&7548.059 & 77.33 $\pm $ 4.46  & 171.56 $\pm $ 7.69 & 32.38 $\pm $ 4.03 & 36.17 $\pm $ 2.91 \\
SWP35403&7551.842 & 65.30 $\pm $ 3.77  & 172.33 $\pm $ 7.72 & 24.95 $\pm $ 3.11 & 39.66 $\pm $ 3.20 \\
SWP35404&7551.901 & 59.71 $\pm $ 3.44  & 151.81 $\pm $ 6.81 & 27.76 $\pm $ 3.45 & 34.32 $\pm $ 2.77 \\
SWP35417&7553.270 & 58.73 $\pm $ 3.39  & 153.76 $\pm $ 6.89 & 32.38 $\pm $ 4.15 & 31.19 $\pm $ 2.51 \\
SWP35428&7554.177 & 50.73 $\pm $ 2.93  & 142.02 $\pm $ 6.37 & 19.38 $\pm $ 2.45 & 37.88 $\pm $ 3.05 \\
SWP35429&7554.232 &\nodata             & \nodata            & \nodata 	& 35.53 $\pm $ 2.86 \\
SWP35457&7556.842 & 51.04 $\pm $ 2.94  & 139.96 $\pm $ 6.27 & 21.38 $\pm $ 2.66 & 31.65 $\pm $ 2.55 \\
SWP35458&7556.908 & 54.85 $\pm $ 3.16  & 139.51 $\pm $ 6.25 & 26.34 $\pm $ 3.28 & 33.77 $\pm $ 2.72 \\
\enddata
\tablenotetext{1}{Continuum fluxes are in units of 10$^{-15}$ erg cm$^{2}
$ s$^{-1}$ \AA$^{-1}$}
\tablenotetext{2}{Emission-line fluxes are in units of 10$^{-13}$ erg cm$^{-2}$ s$^{-1}$}
\end{deluxetable}

%%% TABLE 3 %%%
\begin{deluxetable}{ccc}
\tablewidth{0pt}
\tablecaption{1988 Mg\,{\sc ii}\,$\lambda2798$ Flux Measurements}
\tablehead{
\colhead{Image} &
\colhead{Julian Date} &
\colhead{ } \\
\colhead{Name} &
\colhead{($-2,440,000$)} &
\colhead{Mg\,{\sc ii}\,$\lambda 2798$\tablenotemark{1}} 
}
\startdata
LWP00505&7499.610 &  28.31 $\pm $ 1.71 \\
LWP00506&7499.680 &  29.35 $\pm $ 1.77 \\
LWP14603&7506.100 &  26.01 $\pm $ 1.57 \\
LWP14604&7506.180 &  29.91 $\pm $ 1.81 \\
LWP14660&7513.310 &  25.84 $\pm $ 1.56 \\
LWP14661&7513.380 &  27.25 $\pm $ 1.65 \\
LWP14694&7518.100 &  26.38 $\pm $ 1.60 \\
LWP14695&7518.180 &  26.26 $\pm $ 1.59 \\
LWP14704&7520.070 &  25.26 $\pm $ 1.53 \\
LWP14705&7520.140 &  22.90 $\pm $ 1.38 \\
LWP14730&7524.070 &  22.92 $\pm $ 1.39 \\
LWP14731&7524.140 &  22.64 $\pm $ 1.37 \\
LWP14748&7527.110 &  22.25 $\pm $ 1.35 \\
LWP14773&7532.050 &  21.46 $\pm $ 1.30 \\
LWP14794&7535.990 &  22.51 $\pm $ 1.36 \\
LWP14795&7536.050 &  25.84 $\pm $ 1.56 \\
LWP14814&7540.000 &  25.88 $\pm $ 1.56 \\
LWP14815&7540.070 &  24.40 $\pm $ 1.48 \\
LWP14855&7544.040 &  28.37 $\pm $ 1.72 \\
LWP14888&7548.030 &  28.42 $\pm $ 1.72 \\
LWP14917&7551.870 &  29.07 $\pm $ 1.76 \\
LWP14941&7556.870 &  28.28 $\pm $ 1.71 \\
\enddata
\tablenotetext{1}{Emission-line fluxes are in
units of 10$^{-13}$ erg cm$^{-2}$ s$^{-1}$}
\end{deluxetable}

%%% TABLE 4 %%%
\begin{deluxetable}{lccccc}
\tablewidth{0pt}
\tablecaption{1991 SWP Flux Measurements}
\tablehead{
\colhead{Image } &
\colhead{Julian Date} &
\colhead{ } &
\colhead{ } &
\colhead{ } &
\colhead{ } \\
\colhead{Name} &
\colhead{($-2,440,000$)} &
\colhead{$F_\lambda$(1355\,\AA)\tablenotemark{1}} &
\colhead{C\,{\sc iv}\,$\lambda 1549$\tablenotemark{2}} &
\colhead{He\,{\sc ii}\,$\lambda 1640$\tablenotemark{2}}&
\colhead{C\,{\sc iii}]\,$\lambda 1909$\tablenotemark{2}}
}
\startdata
SWP43044 &8570.007& 245.39 $\pm $ 5.77& 406.54 $\pm $ 13.49& 82.65 $\pm $ 7.99& 64.11 $\pm $ 2.22 \\
SWP43045 &8570.063& 239.55 $\pm $ 5.64& 403.02 $\pm $ 13.37& 76.72 $\pm $ 7.41& 67.28 $\pm $ 2.33 \\
SWP43046 &8570.112& 260.85 $\pm $ 6.14& 420.93 $\pm $ 13.97& 92.43 $\pm $ 8.93& 65.30 $\pm $ 3.36 \\
SWP43052 &8571.005& 254.54 $\pm $ 5.99& 412.64 $\pm $ 13.69& 83.22 $\pm $ 8.04& 62.78 $\pm $ 2.17 \\
SWP43053 &8571.066& 269.34 $\pm $ 6.34& 398.57 $\pm $ 13.22& 85.81 $\pm $ 8.29& \nodata           \\
SWP43117 &8577.048& 203.85 $\pm $ 4.80& 367.67 $\pm $ 12.20& 65.38 $\pm $ 6.32& 62.16 $\pm $ 2.15 \\
SWP43118 &8577.106& 197.79 $\pm $ 4.65& 378.25 $\pm $ 12.55& 76.38 $\pm $ 7.38& 60.35 $\pm $ 2.09 \\
SWP43129 &8578.010& 204.50 $\pm $ 4.81& 377.99 $\pm $ 12.54& 71.34 $\pm $ 6.89& 60.88 $\pm $ 2.11 \\
SWP43130 &8578.070& 199.65 $\pm $ 4.70& 374.75 $\pm $ 12.43& 69.92 $\pm $ 6.76& 61.63 $\pm $ 2.13 \\
SWP43142 &8579.039& 184.62 $\pm $ 4.34& 380.52 $\pm $ 12.62& 64.63 $\pm $ 6.24& 61.61 $\pm $ 2.13 \\
SWP43143 &8579.097& 178.35 $\pm $ 4.20& 366.47 $\pm $ 12.16& 73.22 $\pm $ 7.07& 58.66 $\pm $ 2.03 \\
SWP43155 &8580.221& 142.40 $\pm $ 3.35& 338.72 $\pm $ 11.24& 66.61 $\pm $ 6.44& 57.54 $\pm $ 1.99 \\
SWP43162 &8581.042& 135.03 $\pm $ 3.18& 312.80 $\pm $ 10.38& 59.22 $\pm $ 5.72& 58.29 $\pm $ 2.02 \\
SWP43163 &8581.104& 139.46 $\pm $ 3.28& 342.05 $\pm $ 11.35& 56.62 $\pm $ 5.47& 56.91 $\pm $ 1.97 \\
SWP43172 &8582.041& 118.19 $\pm $ 2.78& 309.46 $\pm $ 10.27& 52.90 $\pm $ 5.11& 59.66 $\pm $ 2.07 \\
SWP43173 &8582.105& 116.39 $\pm $ 2.74& 312.81 $\pm $ 10.38& 50.19 $\pm $ 4.85& \nodata           \\
SWP43190 &8584.037&  94.71 $\pm $ 2.23& 241.56 $\pm $  8.01& 43.53 $\pm $ 4.21& 54.78 $\pm $ 1.90 \\
SWP43191 &8584.107&  95.05 $\pm $ 2.24& 260.56 $\pm $  8.64& 48.69 $\pm $ 4.70& 53.40 $\pm $ 1.85 \\
SWP43209 &8585.011& 125.42 $\pm $ 2.95& 268.44 $\pm $  8.91& 53.17 $\pm $ 5.14& 54.27 $\pm $ 1.88 \\
SWP43210 &8585.085& 122.28 $\pm $ 2.88& 271.76 $\pm $  9.02& 56.85 $\pm $ 5.49& 56.52 $\pm $ 1.96 \\
SWP43218 &8586.016& 101.55 $\pm $ 2.39& 255.73 $\pm $  8.48& 49.47 $\pm $ 4.78& 53.54 $\pm $ 1.85 \\
SWP43219 &8586.092&  95.54 $\pm $ 2.25& 270.95 $\pm $  8.90& 45.61 $\pm $ 4.41& 57.01 $\pm $ 1.97 \\
SWP43228 &8587.026&  99.96 $\pm $ 2.35& 232.37 $\pm $  7.71& 40.45 $\pm $ 3.91& 54.71 $\pm $ 1.89 \\
SWP43229 &8587.093& 102.14 $\pm $ 2.40& 260.29 $\pm $  8.64& 51.99 $\pm $ 5.02& \nodata           \\
SWP43234 &8588.036& 101.84 $\pm $ 2.40& 230.84 $\pm $  7.66& 44.33 $\pm $ 4.28& 50.15 $\pm $ 1.74 \\
SWP43235 &8588.107&  96.60 $\pm $ 2.27& 231.07 $\pm $  7.67& 45.21 $\pm $ 4.37& 53.76 $\pm $ 1.86 \\
SWP43244 &8589.028&  90.57 $\pm $ 2.13& 223.64 $\pm $  7.42& 39.69 $\pm $ 3.83& 51.81 $\pm $ 1.79 \\
SWP43245 &8589.105&  93.37 $\pm $ 2.20& 221.94 $\pm $  7.36& 35.87 $\pm $ 3.47& 50.87 $\pm $ 1.76 \\
SWP43275 &8591.043&  81.99 $\pm $ 1.93& 215.17 $\pm $  7.14& 36.80 $\pm $ 3.56& 49.47 $\pm $ 1.71 \\
SWP43276 &8591.170&  82.17 $\pm $ 1.93& 224.04 $\pm $  7.43& 39.18 $\pm $ 3.79& 45.60 $\pm $ 1.58 \\
SWP43285 &8592.070&  80.64 $\pm $ 1.90& 200.22 $\pm $  6.64& 36.53 $\pm $ 3.53& 47.20 $\pm $ 1.63 \\
SWP43286 &8592.172&  79.79 $\pm $ 1.88& 215.40 $\pm $  7.15& 37.15 $\pm $ 3.59& 49.36 $\pm $ 1.71 \\
SWP43287 &8592.192&  72.84 $\pm $ 1.71& 203.96 $\pm $  6.77& 38.32 $\pm $ 3.70& 47.34 $\pm $ 1.64 \\
SWP43314 &8594.934&  89.00 $\pm $ 2.09& 197.04 $\pm $  6.54& 37.70 $\pm $ 3.64& 47.48 $\pm $ 1.64 \\
SWP43315 &8595.006&  89.68 $\pm $ 9.90& 205.95 $\pm $  6.83& 39.44 $\pm $ 3.81& 47.00 $\pm $ 1.63 \\
SWP43323 &8596.061&  83.24 $\pm $ 1.96& 174.03 $\pm $  5.77& 34.55 $\pm $ 3.34& 43.31 $\pm $ 1.50 \\
SWP43324 &8596.138&  82.76 $\pm $ 1.95& 193.47 $\pm $  6.42& 35.98 $\pm $ 3.48& 46.14 $\pm $ 1.60 \\
SWP43325 &8596.191&  79.34 $\pm $ 1.87& 200.13 $\pm $  6.64& 29.23 $\pm $ 2.82& 41.69 $\pm $ 1.44 \\
SWP43333 &8597.099& 102.94 $\pm $ 2.42& 182.50 $\pm $  6.06& 36.23 $\pm $ 3.50& 44.57 $\pm $ 1.54 \\
SWP43334 &8597.170& 104.69 $\pm $ 2.46& 196.99 $\pm $  6.54& 39.83 $\pm $ 3.85& \nodata           \\
SWP43340 &8597.936& 164.78 $\pm $ 3.88& 201.26 $\pm $  6.68& 47.49 $\pm $ 4.59& 51.18 $\pm $ 1.77 \\
SWP43341 &8598.014& 169.04 $\pm $ 3.98& 210.27 $\pm $  6.97& 48.12 $\pm $ 4.65& 53.99 $\pm $ 1.87 \\
SWP43373 &8604.141& 198.04 $\pm $ 4.66& 310.46 $\pm $ 10.30& 54.47 $\pm $ 5.26& 58.17 $\pm $ 2.01 \\
SWP43392 &8606.188& 182.24 $\pm $ 4.29& 324.30 $\pm $ 10.76& 65.95 $\pm $ 6.37& 55.72 $\pm $ 1.93 \\
\enddata
\tablenotetext{1}{Continuum values are in units of 10$^{-15}$ 
erg cm$^{-2}$ s$^{-1}$ \AA$^{-1}$ .}
\tablenotetext{2}{Emission line fluxes are in units of 10$^{-13}$ erg
 cm$^{-2}$ s$^{-1}$.}
\end{deluxetable}

%%% TABLE 5 %%%
\begin{deluxetable}{ccc}
\tablewidth{0pt}
\tablecaption{1991 Mg\,{\sc ii}\,$\lambda2798$ Flux Measurements}
\tablehead{
\colhead{Image} &
\colhead{Julian Date} &
\colhead{ } \\
\colhead{Name} &
\colhead{($-2,440,000$)} &
\colhead{Mg\,{\sc ii}\,$\lambda 2798$\tablenotemark{1}} 
}
\startdata
LWP21671 &8570.530&  54.45 $\pm $ 1.56 \\
LWP21681 &8571.520&  58.60 $\pm $ 1.70 \\
LWP21750 &8577.500&  57.02 $\pm $ 1.65 \\
LWP21751 &8577.570&  56.87 $\pm $ 1.64 \\
LWP21752 &8577.620&  60.61 $\pm $ 1.75 \\
LWP21764 &8578.480&  58.65 $\pm $ 1.70 \\
LWP21765 &8578.530&  57.92 $\pm $ 1.68 \\
LWP21774 &8579.560&  56.47 $\pm $ 1.63 \\
LWP21775 &8579.610&  57.10 $\pm $ 1.65 \\
LWP21785 &8580.750&  54.26 $\pm $ 1.57 \\
LWP21791 &8581.570&  54.56 $\pm $ 1.58 \\
LWP21792 &8581.620&  53.27 $\pm $ 1.54 \\
LWP21798 &8582.570&  52.71 $\pm $ 1.52 \\
LWP21827 &8584.560&  51.87 $\pm $ 1.50 \\
LWP21835 &8585.540&  48.67 $\pm $ 1.41 \\
LWP21836 &8585.610&  50.59 $\pm $ 1.46 \\
LWP21845 &8586.550&  50.20 $\pm $ 1.45 \\
LWP21854 &8587.550&  45.81 $\pm $ 1.33 \\
LWP21855 &8587.610&  44.98 $\pm $ 1.30 \\
LWP21863 &8588.560&  52.89 $\pm $ 1.53 \\
LWP21872 &8589.560&  49.81 $\pm $ 1.44 \\
LWP21873 &8589.630&  48.52 $\pm $ 1.40 \\
LWP21893 &8591.460&  46.38 $\pm $ 1.34 \\
LWP21894 &8591.570&  43.59 $\pm $ 1.26 \\
LWP21904 &8592.600&  46.45 $\pm $ 1.34 \\
LWP21905 &8592.660&  49.40 $\pm $ 1.43 \\
LWP21930 &8595.460&  47.18 $\pm $ 1.36 \\
LWP21931 &8595.530&  46.71 $\pm $ 1.35 \\
LWP21943 &8596.590&  43.65 $\pm $ 1.26 \\
LWP21956 &8597.630&  42.06 $\pm $ 1.22 \\
LWP21965 &8598.460&  49.16 $\pm $ 1.42 \\
\enddata
\tablenotetext{1}{Emission-line fluxes are in
units of 10$^{-13}$ erg cm$^{-2}$ s$^{-1}$}
\end{deluxetable}

%%% TABLE 6 %%%
\begin{deluxetable}{lccccccc}
\tablewidth{0pt}
\tablecaption{Light Curve Statistics}
\tablehead{
\colhead{ } &
\colhead{ } &
\multicolumn{2}{c}{Sampling} &
\colhead{ } &
\colhead{Mean} \\
\colhead{Time} &
\colhead{ } &
\multicolumn{2}{c}{Interval (days)} &
\colhead{Mean} &
\colhead{Fractional} \\
\colhead{Series } &
\colhead{$N$} &
\colhead{$\langle T \rangle$} &
\colhead{$T_{\rm median}$} &
\colhead{Flux\tablenotemark{1}} &
\colhead{Error} &
\colhead{$F_{\rm var}$} &
\colhead{$R_{\rm max}$} \\
\colhead{(1)} &
\colhead{(2)} &
\colhead{(3)} &
\colhead{(4)} &
\colhead{(5)} &
\colhead{(6)} &
\colhead{(7)} &
\colhead{(8)} 
}
\startdata
1988 1355\,\AA &
18 & 3.4 & 3.8 & $58.1 \pm 17.3$ & 0.046 & 0.293 & $2.65\pm0.15$ \\
1988 C\,{\sc iv}&
19 & 3.3 & 3.6 & $154.2\pm35.6$  & 0.036 & 0.228 & $2.28\pm0.13$ \\
1988 He\,{\sc ii}& 
19 & 3.3 & 3.5 & $27.7\pm10.7$   & 0.100 & 0.369 & $3.89\pm0.60$ \\
1988 C\,{\sc iii}]&
19 & 3.3 & 3.6 & $34.2\pm2.4$    & 0.066 & 0.018 & $1.29\pm0.10$ \\
1988 Mg\,{\sc ii} &
14 & 4.4 & 4.0 & $25.9\pm2.6$    & 0.050 & 0.088 & $1.36\pm0.12$ \\
1991 1355\,\AA &
22 & 1.7 & 1.0 & $140.0\pm56.1$  & 0.039 & 0.390 & $3.38\pm0.17$ \\
1991 C\,{\sc iv}&
22 & 1.7 & 1.0 & $283.6\pm73.7$  & 0.024 & 0.259 & $2.17\pm0.06$ \\
1991 He\,{\sc ii}& 
22 & 1.7 & 1.0 & $53.6\pm15.3$   & 0.070 & 0.275 & $2.58\pm0.23$ \\
1991 C\,{\sc iii}]&
22 & 1.7 & 1.0 & $54.8\pm6.0$    & 0.027 & 0.106 & $1.50\pm0.043$ \\
1991 Mg\,{\sc ii} &
20 & 1.5 & 1.0 & $51.0\pm5.0$    & 0.025 & 0.094 & $1.39\pm0.06$ \\
\enddata
\tablenotetext{1}{Continuum and emission-line fluxes are in the same units
used in Tables 2--5.}
\end{deluxetable}

%%% TABLE 7 %%%
\begin{deluxetable}{lccc}
\tablewidth{0pt}
\tablecaption{Cross-Correlation Results}
\tablehead{
\colhead{ } &
\colhead{ } &
\colhead{$\tau_{\rm cent}$\tablenotemark{1}} &
\colhead{$\tau_{\rm peak}$\tablenotemark{1}} \\
\colhead{Line} &
\colhead{$r_{\rm max}$} &
\colhead{(days)} &
\colhead{(days)} \\
\colhead{(1)} &
\colhead{(2)} &
\colhead{(3)} &
\colhead{(4)} 
}
\startdata
1988 C\,{\sc iv}$\,\lambda1549$&
$0.882\pm0.055$ & $3.44^{+1.42}_{-1.24}$ & $3.1^{+2.0}_{-1.1}$ \\
1988 He\,{\sc ii}\,$\lambda1640$& 
$0.819\pm0.078$ & $3.47^{+1.97}_{-1.61}$ & $3.6^{+1.9}_{-3.5}$ \\
1988 C\,{\sc iii}]\,$\lambda1909$&
$0.709\pm0.107$ & $6.90^{+4.58}_{-3.83}$ & $6.5^{+4.9}_{-3.9}$ \\
1988 Mg\,{\sc ii}\,$\lambda2798$&
$0.883\pm0.064$ & $6.83^{+1.74}_{-2.10}$ & $6.6^{+2.2}_{-1.6}$ \\
1991 C\,{\sc iv}\,$\lambda1549$&
$0.965\pm0.018$ & $3.28^{+0.83}_{-0.91}$ & $3.3^{+0.5}_{-0.8}$ \\
1991 He\,{\sc ii}\,$\lambda1640$& 
$0.933\pm0.034$ & $2.60^{+1.10}_{-1.21}$ & $1.5^{+1.2}_{-1.0}$ \\
1991 C\,{\sc iii}]\,$\lambda1909$&
$0.884\pm0.048$ & $3.45^{+1.52}_{-1.22}$ & $3.6^{+2.0}_{-2.2}$ \\
1991 Mg\,{\sc ii}\,$\lambda2798$ &
$0.925\pm0.177$ & $5.35^{+1.87}_{-1.77}$ & $4.6^{+2.8}_{-2.5}$ \\
\enddata
\tablenotetext{1}{Lags are in the observed frame.}
\end{deluxetable}

%%% TABLE 8 %%%

\begin{deluxetable}{lll}
\tablewidth{0pt}
\tablecaption{Emission Line Widths}
\tablehead{
\colhead{ } &
\colhead{$\sigma_{\rm line}$} &
\colhead{FWHM} \\
\colhead{Line} &
\colhead{(km\ s$^{-1}$)} &
\colhead{(km\ s$^{-1}$)} \\
\colhead{(1)} &
\colhead{(2)} &
\colhead{(3)}
}
\startdata
1988 C\,{\sc iv}\,$\lambda1549$\tablenotemark{1}   
            & $5698\pm245$ & $6697\pm543$   \\
1988 He\,{\sc ii}\,$\lambda1640$\tablenotemark{2}  
            & $5013\pm323 $ & $5356\pm1270$ \\
1988 C\,{\sc iii}]\,$\lambda1909$ 
            & $2553\pm 307$ & $2646\pm745$ \\
1988 Mg\,{\sc ii}\,$\lambda2798$  
            & $2581\pm179$ & $4823\pm1105$  \\
1991 C\,{\sc iv}\,$\lambda1549$\tablenotemark{1}
            & $5140\pm113$ & $4858\pm149$ \\ 
1991 He\,{\sc ii}\,$\lambda1640$\tablenotemark{2}  
            & $4530\pm92$  & $4597\pm659$ \\ 
1991 C\,{\sc iii}]\,$\lambda1909$ 
            & $2817\pm81 $ & $6997\pm1366$ \\
1991 Mg\,{\sc ii}\,$\lambda2798$ 
            & $2721\pm141$ & $6458\pm1850$ \\
\enddata
\tablenotetext{1}{Based on the shortward side of the line only.}
\tablenotetext{2}{Based on the longward side of the line only.}
\end{deluxetable}

\end{document}